\documentclass[%
 amsmath,amssymb,
 reprint,%
]{revtex4-1}
\usepackage{siunitx}
\usepackage{color}
\usepackage{graphicx}
\usepackage{dcolumn}
\usepackage{bm}

\def\clr{\color{red}}
\begin{document}

\title{High-resolution imaging of Rydberg atoms in optical lattices using an aspheric-lens objective in vacuum}

\author{Chuyang Shen}
\author{Cheng Chen}
\author{Xiao-Ling Wu}
\author{Shen Dong}
\author{Yue Cui}
\affiliation{
State Key Laboratory of Low-Dimensional Quantum Physics, Department of Physics, Tsinghua University, Beijing 100084, China
}%
\author{Li You}
\email{lyou@mail.tsinghua.edu.cn}
\author{Meng Khoon Tey}
\email{mengkhoon\_tey@mail.tsinghua.edu.cn}
\affiliation{
State Key Laboratory of Low-Dimensional Quantum Physics, Department of Physics, Tsinghua University, Beijing 100084, China
}%
\affiliation{%
Frontier Science Center for Quantum Information, Beijing 100084, China
}%

\date{\today}

\def\clr{\color{red}}

\begin{abstract}

We present a high-resolution, simple and versatile system for imaging ultracold Rydberg atoms in optical lattices. The imaging objective is a single aspheric lens (with a working distance of 20.6\,mm  and a numerical aperture (NA) of 0.51) placed inside the vacuum chamber. Adopting a large-working-distance lens leaves room for electrodes and electrostatic shields to control electric fields around Rydberg atoms. With this setup, we achieve an Rayleigh resolution of 1.10\,\si{\micro m} or $1.41\lambda$ ($\lambda=780$\,nm), limited by the NA of the aspheric lens. For systems of highly excited Rydberg states with blockade radii greater than a few \si{\micro m}, the resolution achieved is sufficient for studying many physical processes of interest.
\end{abstract}

\maketitle

\section{\label{sec:INTRODUCTION}INTRODUCTION}

High-resolution in-situ quantum microscope for ultracold atoms in optical lattices~\cite{bakr2009quantum,gemelke2009situ,sherson2010single,parsons2015site,cheuk2015quantum,haller2015single,miranda2015site,picken2017single,yamamoto2017site,gempel2019adaptable,yang2020observation,Nelson2007} allows observation and control of ultracold atoms down to the single particle level. Such technology, accompanied with the capabilities to control contact interaction between ultracold atoms using technique like magnetic Feshbach resonance~\cite{Chin2010FRreview, Cui2018Broad} and to establish long-range interaction by using highly excited Rydberg states~\cite{saffman2010quantum}, makes cold atom experiment a rich playground for studying exotic many-body states not accessible by other physical systems and for performing quantum simulations of various physical models of importance~\cite{bakr2010probing,mazurenko2017cold,keesling2019quantum,bernien2017probing,kim2018detailed,Takei2016,Browaeys2020}.

At present, most quantum microscope setups adopt custom-made objectives placed outside of the vacuum chamber to collect fluorescence light from atoms~\cite{bakr2009quantum,gemelke2009situ,sherson2010single,parsons2015site,cheuk2015quantum,haller2015single,miranda2015site,picken2017single,yamamoto2017site,gempel2019adaptable}. These objectives are expensive and require strict conditions for the thickness and the flatness of the viewport glass to work as designed \cite{Xie2018Development}. Besides, large NA objectives typically need to be placed very close to the atoms, making it difficult to insert components like electrodes and electrostatic shields around the atoms. Such components are important if one is to work with Rydberg atoms of a large principal quantum number, which are sensitive to stray electric fields. Furthermore, since Rydberg atoms feature a large interaction range (dipole-blockade radius) typically larger than a few microns~\cite{saffman2010quantum}, many physical processes of interests, such as quantum transport of Rydberg excitons with large principal quantum numbers~\cite{gunter2013observing,yang2019quantum}, Rydberg macrodimers~\cite{hollerith2019quantum}, and ordering in Rydberg many-body systems~\cite{schaus2012observation}, can be observable without achieving sub-micron imaging resolution.

This work introduces a simple setup for imaging and controlling ultracold Rydberg atoms in optical lattices. Our design makes use of a single aspheric lens placed inside the vacuum chamber as the main imaging objective. This lens features a large working distance of 20.6\,mm, which provides sufficient space for inserting electrodes and electrostatic shields around the Rydberg atoms. Using this setup, we achieve an Rayleigh resolution of 1.10\,\si{\micro m} or $1.41\lambda$ ($\lambda=780$\,nm), limited by the NA of the aspheric lens but sufficient for studying many physical problems based on the long-range Rydberg interaction~\cite{keesling2019quantum, bernien2017probing,kim2018detailed,Takei2016,Browaeys2020,Tong2004, Heidemann2007,schaus2012observation,Faoro2015,Zeng2017,Cubel2005,gunter2013observing}. Our design allows for a simpler quantum microscope on Rydberg atoms.

\section{\label{sec:EXPERIMENTAL_SETUP}EXPERIMENTAL SETUP}

\begin{figure}
  \centering
  \includegraphics[width=\columnwidth]{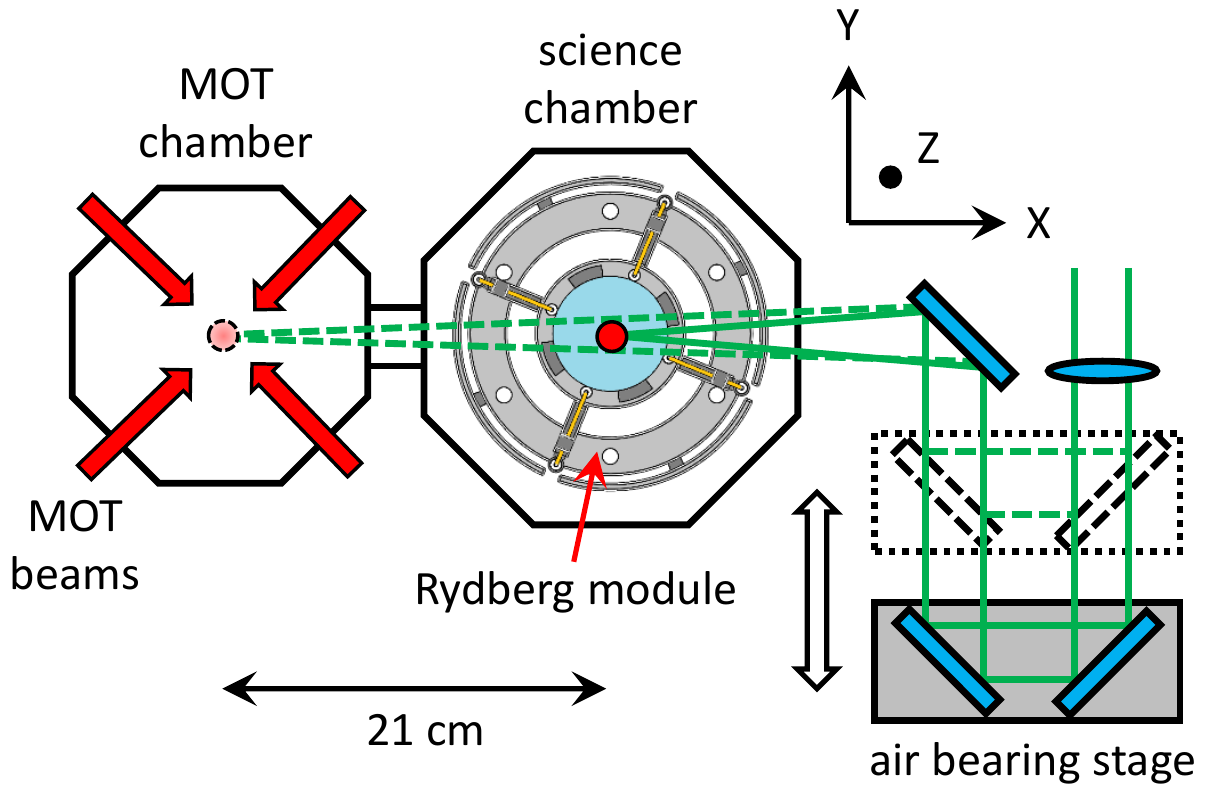}\\
  \caption{Main vacuum chamber composed of a MOT compartment and a science compartment. $^{87}$Rb atoms are captured and cooled to 12\,\si{\micro K} in the MOT chamber, and then transported to the science chamber using a moving optical dipole trap. The science chamber contains a module built for high-resolution imaging and for controlling electric field around the Rydberg atoms.}\label{fig:vacuum}
\end{figure}

\begin{figure*}[ht]
\centering
\includegraphics[width=17cm]{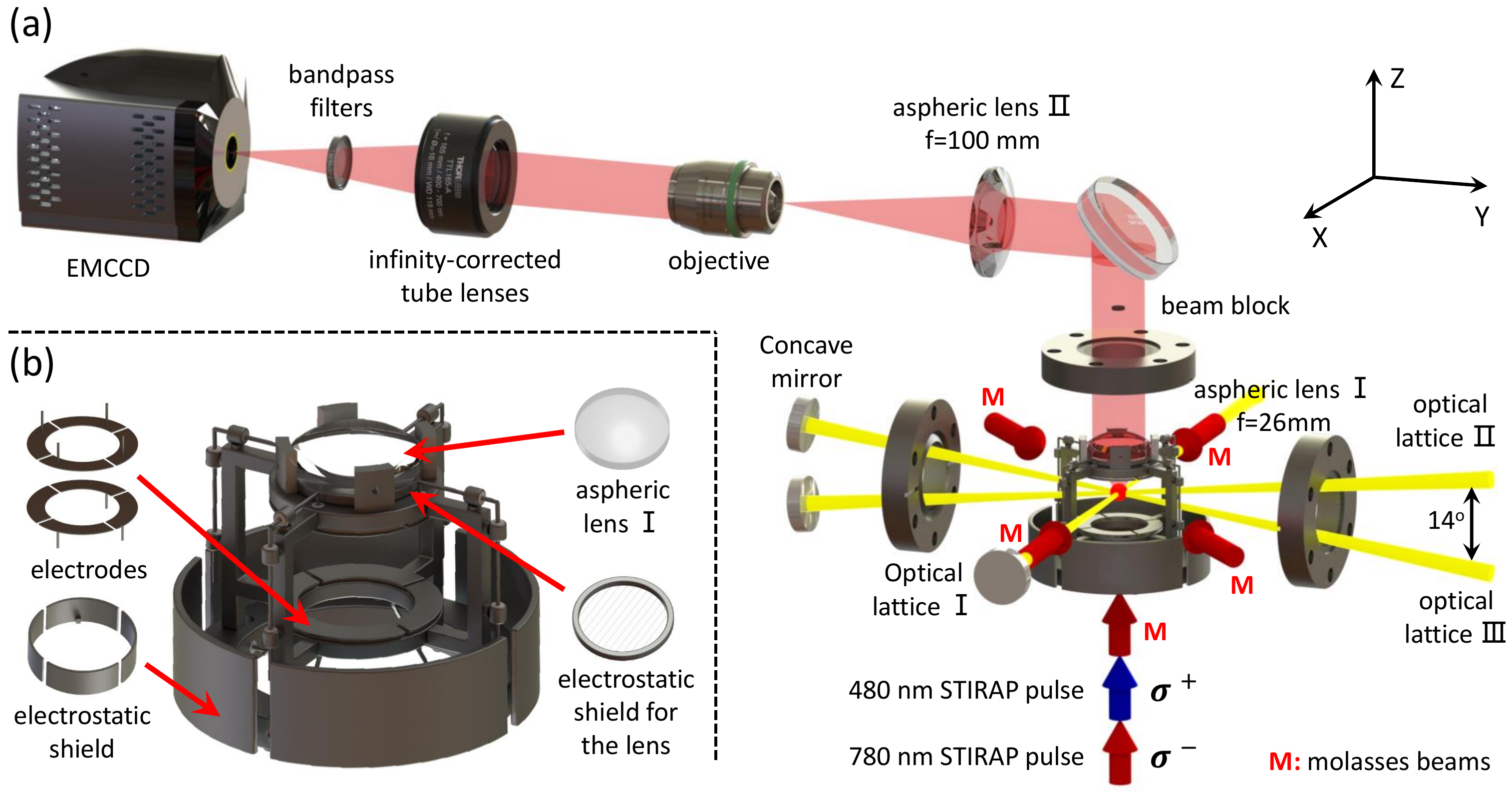}
\caption{Experimental setup for production, control, and imaging of Rydberg atoms. (a) 3D illustration of the fluorescence imaging setup, and light beams for trapping and exciting Rydberg atoms. (b) The ``Rydberg module'' packs the aspheric lens~\uppercase\expandafter{\romannumeral1} for collecting fluorescence light, eight electrodes and electrostatic shields for controlling the electric field around the Rydberg atoms.}\label{fig:imaging_system}
\end{figure*}

Our ultrahigh vacuum chamber (Fig.~\ref{fig:vacuum}) has two main compartments: a magneto-optical-trap (MOT) chamber and a science chamber. $^{87}$Rb atoms are first cooled and captured in the MOT chamber. After the MOT and molasses cooling stages, atoms are optically pumped into the $|F=1,m_F=-1\rangle$ hyperfine ground state and trapped by a magnetic quadrupole field. They are subsequently cooled by evaporation down to 12\,\si{\micro K} using microwave radiation, and then transferred from the magnetic trap into the focus of a 1064-nm 10-W light beam with a waist of about 31\,\si{\micro m}. Finally, the optically trapped atoms are transported from the MOT chamber to the science chamber by moving the focus of this light beam over 21\,cm using an air bearing stage (Fig.~\ref{fig:vacuum}). In the science chamber, we further cool the atoms to sub \si{\micro K} by weakening the optical trap and then transfer the atoms into a three-dimensional optical lattice. Excitation to Rydberg states are conducted hereafter.

The optical lattices are composed of three 1064-nm standing waves, namely, lattice~\uppercase\expandafter{\romannumeral1}, lattice~\uppercase\expandafter{\romannumeral2}, and lattice~\uppercase\expandafter{\romannumeral3} as shown in Fig.~\ref{fig:imaging_system}(a). Lattice~\uppercase\expandafter{\romannumeral1} is along the horizontal X direction. Lattice~\uppercase\expandafter{\romannumeral2} and lattice~\uppercase\expandafter{\romannumeral3} are on the Y-Z plane and are at angles of $\pm7$
degrees with respect to the Y-axis. To save space, each of the lattices is formed by reflection using a single concave mirror, such that the foci before and after the mirror overlap at the atoms. The waists of lattices ~\uppercase\expandafter{\romannumeral1}, \uppercase\expandafter{\romannumeral2}, and \uppercase\expandafter{\romannumeral3} are about 68\,\si{\micro m}, 78\,\si{\micro m}, and 78\,\si{\micro m}, respectively. To avoid interference-induced fluctuations on the lattice depths, the light frequencies of three lattices are offset by at least 220\,MHz with respect to each other. At a power of 10\,W in each lattice beam, the depth of our 3D optical lattice is about $k_B\cdot0.62$\,mK ($k_B$ being the Boltzmann constant). As lattices~\uppercase\expandafter{\romannumeral2} and \uppercase\expandafter{\romannumeral3} are not perpendicular to each other, our lattice is not a cubic lattice. It is approximately a square lattice in the X-Y plane with a spacing of 532\,nm. Each of such layers are spaced by 2.2\,\si{\micro m} along the Z-axis, with adjacent layers displaced by 266\,nm along the Y-direction. Adopting a larger spacing along the Z-direction makes it easier to selectively prepare a layer of atomic ensemble (in the X-Y plane) with the help of a gradient magnetic field along the Z-direction. Besides, atoms out of the focal plane would contribute less to the recorded image since our imaging system is constructed along the Z-direction.

We image atoms in the optical lattices by detecting fluorescence light of atoms on a EMCCD camera. A single aspheric lens (aspheric lens I in Fig.~\ref{fig:imaging_system}(a)) with a NA of 0.51 (focal length = 26\,mm, working distance = 20.6\,mm, Thorlabs AL3026-B), placed inside the science chamber, collects radiations from the atoms. As the atoms are placed at/near the focus of the lens, the fluorescence light is collimated by the lens before passing through the viewport (Fig.~\ref{fig:imaging_system}(a)). This arrangement voids the strict demands on the thickness and flatness of viewport glass, in comparison to setups with objectives placed outside of the vacuum chamber~\cite{Xie2018Development}. Real images of the atoms are created by focusing the collimated radiation using a second aspheric lens with a focal length of 100\,mm outside of the vacuum chamber, resulting in a magnification of 3.85. These atom images are further magnified by 20 times using a microscope composed of an objective from ZEISS (Objective EC Epiplan-Neofluar 20$\times$-0.50 BD DIC M27) and an infinity-corrected tube lens from Thorlabs (TTL165-A). The ultimate images are recorded using an 512$\times$512-pixel EMCCD camera (Andor iXon3) with a pixel area of 16\,\si{\micro\metre}$\times$16\,\si{\micro\metre}. Overall, the size of one pixel on the camera corresponds to 0.208\,\si{\micro\metre}$\times$ 0.208\,\si{\micro\metre} in the object plane. To prevent unwanted stray light from contaminating the signal from the atoms, the optical path of the imaging system is carefully shielded wherever possible and a bandpass filter with a central wavelength of 780\,nm and a bandwidth of 10\,nm is placed in front of the camera.

We combine the first objective (aspheric lens I) in our imaging system with an octopole-electrode structure and a set of electrostatic shields to form a compact module as shown in Fig.~\ref{fig:imaging_system}(b). The octopole-electrode structure are constructed from eight 0.5-mm thick stainless steel plates. By controlling the voltage on each plates individually, one can generate many electric field configurations, including constant fields in arbitrary directions, gradient fields, quadrupole fields, etc~\cite{low2006versatile}. To shield possible electric fields from charges accumulated on the dielectric aspheric lens, a metal mesh is placed between the lens and the atoms as shown in Fig.~\ref{fig:imaging_system}(b).

To realize in-situ fluorescence imaging of atoms in optical lattices, five optical molasses beams (see Fig.~\ref{fig:imaging_system}(a)) are used to excite and at the same time to cool the atoms to prevent site-hopping transport~\cite{weitenberg2011single}. As the molasses beam along the Z-direction points towards the EMCCD camera and produces significant background noise, a small beam block with a diameter of 4.5\,mm is placed between aspheric lens~\uppercase\expandafter{\romannumeral1} and aspheric lens~\uppercase\expandafter{\romannumeral2} to fully block this beam, at the expense of reducing the signal by about 2.5\%.

It is crucial to make sure that all lenses are coaxial and the atoms are placed near the focus of aspheric lens, in order to achieve near diffraction limited imaging resolution. For coaxial alignment, we adopt a collimated 780-nm light beam which coincides with its back reflection from a mirror, and place the optical components one by one. Each time, we optimize the overlap between the incident and reflected beams as much as we can. To place the atoms on the optical axis of our imaging system, we shine a collimated light beam, which is on resonance with the atoms, on aspheric lens I along the optical axis, and use this beam to blow away the atoms. We move the position of the optical dipole trap until the effects of the beam on the atoms are maximized. Finally, to make sure the atoms are near the ideal focal plane of aspheric lens I, we adjust the position of dipole trap along the $Z$-direction (and correspondingly the position of the camera) until we obtain the sharpest fluorescence image.

\begin{figure}
  \centering
  \includegraphics[width=\columnwidth]{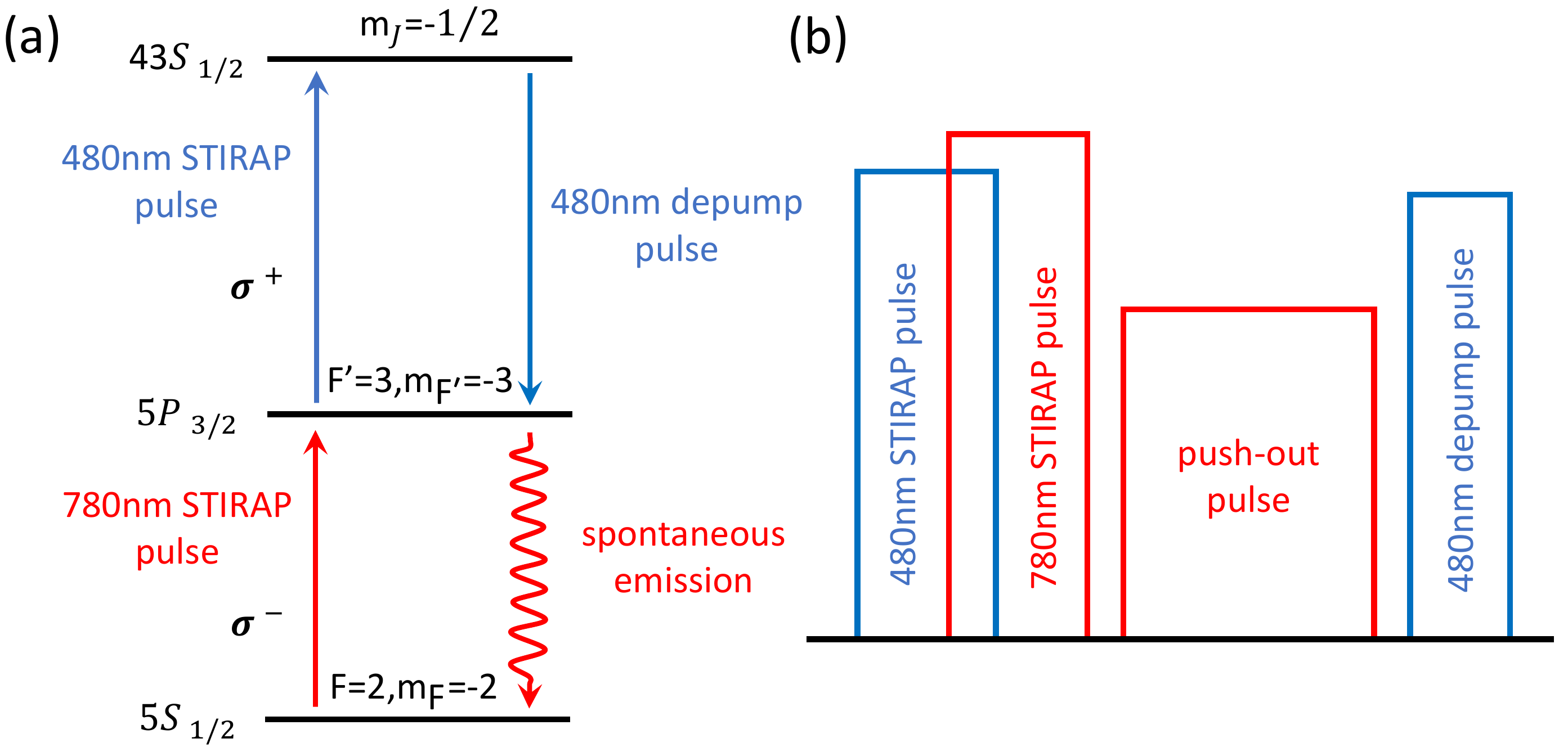}\\
  \caption{Rydberg excitation scheme. (a) Relevant energy levels of a $^{87}$Rb atom for current experiment. (b) Light-pulse sequence used for preparing Rydberg atoms and subsequent detection.}\label{fig:STIRAP}
\end{figure}

To prepare Rydberg atoms, we use stimulated Raman adiabatic passage~\cite{bergmann1998coherent} (STIRAP) with two lasers at 780\,nm and 480\,nm to excite ultracold $^{87}$Rb atoms from the ground $5S\left| {F = 2,m_F =  - 2} \right\rangle$ to the $43S\left|J=1/2,m_J=-1/2\right\rangle$ state (Fig.~\ref{fig:STIRAP}(a)). The 780-nm (480-nm) light propagates along a bias magnetic field of 4.75\,G in the +Z-direction and is $\sigma^{-(+)}$-polarized (Fig.~\ref{fig:imaging_system}(a)). After a counter-intuitive STIRAP pulse sequence (a 480-nm light pulse followed by a 780-nm light pulse), a 780-nm push-out pulse is used to blow away all atoms remaining in the $5S$ ground state (as illustrated in Fig.~\ref{fig:STIRAP}(b)). At last, atoms in the Rydberg states are pumped back to the $5S$ ground state using a 480-nm light pulse in preparation for final fluorescence imaging. In the current experiment, the STIRAP, push-out, and depumping pulses take 5\,\si{\micro}s, 8\,\si{\micro}s, and 2\,\si{\micro}s, respectively. In order to ensure that the atoms remain on the same lattice site (which requires atom displacement smaller than half the lattice spacing) during the whole 15\,\si{\micro s}, the temperature of the gas should be less than $\sim$100\,nK~\cite{recoilmovement}. This rather stringent requirement on the atom temperature can be relaxed once we can prepare a 2D sample overlapping with the objective's focal plane, since the latter would greatly reduce the duration of the push-out pulse.

Right after the 480-nm deexcitation pulse, the depth of the 3D lattice is set to $k_B\cdot0.3$\,mK and then ramped linearly to $k_B\cdot0.6$\,mK in 2\,ms, before the molasses beams for imaging and cooling are turned on. To estimate atom loss and site-hopping probabilities during the fluorescence imaging, we take two consecutive fluorescence images of the same sample. The acquisition duration of each fluorescence image is 500\,\si{ms}. In between of the two acquisitions, the molasses beams are turned off for 1000\,\si{ms}. Comparison between the two images show that about 9\% of the atoms hop over the sites and about 4\% of the atoms are lost. These numbers are expected to improve with more optimizations.

\section{\label{sec:RESULTS}Imaging resolution}

\begin{figure}[t]
  \centering
  \includegraphics[width=0.9\columnwidth]{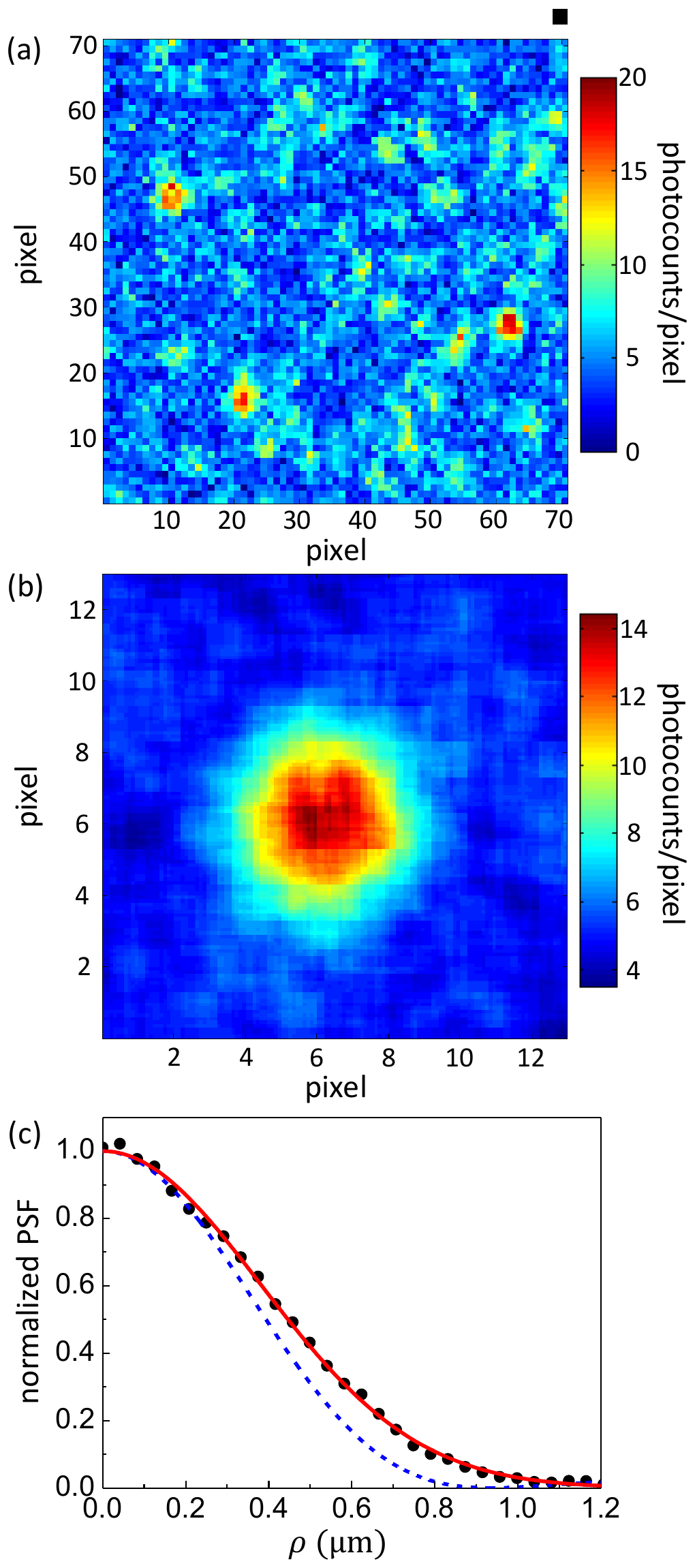}
  \caption{(a) A fluorescence image of (deexcited) Rydberg atoms in optical lattice. One pixel on the EMCCD camera corresponds to an area of 208\,nm $\times$ 208\,nm in the object plane. The black solid square on the top right corner represents the size of a 532\,nm by 532\,nm lattice cell. (b) The point spread function (PSF) of 25 aligned and averaged isolated bright spots. (c) The radial PSF. The black solid dots are experimental data from averaging circularly-symmetric PSF in (b) over azimuthal angles. A small offset due to background noise is removed and the signal is normalized to the maximum for ease of comparison. The red solid line represents a Gaussian fit to the experimental data. The blue dashed line represents the ideal PSF for a lens with NA = 0.51. }\label{fig:Rydberg_atoms}
\end{figure}

Figure~\ref{fig:Rydberg_atoms}(a) shows an example of the fluorescence image of (deexcited) Rydberg atoms. The imaging time of this picture is 1000\,ms. On average, about 374 photons are detected by the EMCCD camera for every atom during this period. Due to Rydberg blockade, no two atoms can be excited within a dipole-blockade radius of about 5\,\si{\micro m}. Therefore, each of the bright spots in the Fig.~\ref{fig:Rydberg_atoms}(a) most likely represents a single Rydberg atom. However, as we do not prepare the atoms in a single layer (in the X-Y plane) at the current stage of our experiment, atoms out of the focal plane can also be observed. As a result, the distance between some atoms may appear to be smaller than the blockade radius.

To quantify the resolution of our imaging system, we align and average the signal from 25 bright and isolated atoms to obtain the point spread function (PSF) of the imaging system (Fig.~\ref{fig:Rydberg_atoms}(b)). The circularly symmetric PSF is further averaged over the azimuthal angle to give a PSF depending only on the radial coordinate $\rho$, which further increases the SNR. After subtracting a small offset originating from background noise from this 1D PSF and normalize its maximum value to 1, the resulting PSF is plotted as black solid dots in Fig.~\ref{fig:Rydberg_atoms}(c). Fitting the normalized PSF with a Gaussian function (red solid line) gives a FWHM of $0.892$\,\si{\micro m}, corresponding to a Rayleigh resolution of 1.10\,\si{\micro m}. The latter value is slightly larger than the theoretical limit of $r_\mathrm{Airy} = 1.22\lambda/\left( 2\mathrm{NA} \right) = 0.933$\,\si{\micro m} for the aspheric lens we use. Nevertheless, for $^{87}$Rb atoms in the $43S$ state with blockade radius of about 5\,\si{\micro m}, the resolution of our imaging system fully meets the needs of our future experiments.

\section{\label{sec:CONCLUSION}CONCLUSION}

In conclusion, we use an off-the-shelf aspheric lens to build a high-resolution imaging system for probing Rydberg atoms in optical lattices. The large working distance of the aspheric lens in the vacuum chamber provides sufficient space around atoms to put electrodes and electrostatic shields for controlling the electric field around Rydberg atoms. Placing the lens inside the vacuum chamber relaxes the requirements on the flatness and thickness of the viewport. We achieve an imaging resolution of 1.10\,\si{\micro m}, limited mainly by the NA of the lens we choose. A higher resolution can be achieved by using an aspheric lens with a larger NA. Our design provides an practical alternative to quantum microscope using customized objectives placed outside the vacuum chamber, and is particular useful if one is to work with Rydberg states which are very close to the disassociation threshold. The setup can be, for examples, used to study quantum transport of Rydberg excitons with large principal quantum numbers~\cite{gunter2013observing,yang2019quantum}, Rydberg macrodimers~\cite{hollerith2019quantum}, and ordering in Rydberg many-body systems~\cite{schaus2012observation}.

\begin{acknowledgments}
This work is supported by the National Key R\&D Program of China (Grant No. 2018YFA0306503 and No. 2018YFA0306504) and by the NSFC (Grant No. 91636213, No. 91736311, No. 91836302, and No. 11654001).
\end{acknowledgments}


\begin{thebibliography}{36}
\bibitem [{rec()}]{recoilmovement}%
  \BibitemOpen
  \href@noop {} {}\bibinfo {note} {In 15\,\si{\micro s}, recoils from the
  480-nm and 780-nm photons would move the atoms over 230\,nm, much smaller
  than the 2.2-\si{\micro m} lattice spacing in the $Z$-direction.}\BibitemShut
  {Stop}%
\end{thebibliography}
%

\end{document}